\documentclass[%
 reprint,
 amsmath,amssymb,
 aps,
]{revtex4-2}

\usepackage{graphicx}
\usepackage{dcolumn}
\usepackage{bm}
\usepackage{amsmath}
\usepackage{enumitem}
\usepackage{color}

\begin{document}

\preprint{APS/123-QED}

\title{Evidence of variable range hopping in the Zintl phase EuIn$_2$P$_2$}

\author{Tomasz Toli\'nski}
\email{tomtol@ifmpan.poznan.pl}
\affiliation{Institute of Molecular Physics, Polish Academy of Sciences, M. Smoluchowskiego 17, 60-179 Pozna\'n, Poland}

\author{Qurat Ul Ain}
\affiliation{Institute of Molecular Physics, Polish Academy of Sciences, M. Smoluchowskiego 17, 60-179 Pozna\'n, Poland}

\author{Karol Synoradzki}
\affiliation{Institute of Molecular Physics, Polish Academy of Sciences, M. Smoluchowskiego 17, 60-179 Pozna\'n, Poland}

\author{Andrzej \L{}api\'nski}
\affiliation{Institute of Molecular Physics, Polish Academy of Sciences, M. Smoluchowskiego 17, 60-179 Pozna\'n, Poland}

\author{Sylwia Zi\c{e}ba}
\affiliation{Institute of Molecular Physics, Polish Academy of Sciences, M. Smoluchowskiego 17, 60-179 Pozna\'n, Poland}

\author{Tetiana Romanova}
\affiliation{Institute of Low Temperature and Structure Research, Polish Academy of Sciences, Ok\'olna 2, 50-422 Wroc{\l}aw, Poland}

\author{Karan Singh}
\affiliation{Institute of Low Temperature and Structure Research, Polish Academy of Sciences, Ok\'olna 2, 50-422 Wroc{\l}aw, Poland}

\author{Orest Pavlosiuk}
\affiliation{Institute of Low Temperature and Structure Research, Polish Academy of Sciences, Ok\'olna 2, 50-422 Wroc{\l}aw, Poland}

\author{Piotr Wi\'sniewski}
\affiliation{Institute of Low Temperature and Structure Research, Polish Academy of Sciences, Ok\'olna 2, 50-422 Wroc{\l}aw, Poland}

\author{Dariusz Kaczorowski}
\affiliation{Institute of Low Temperature and Structure Research, Polish Academy of Sciences, Ok\'olna 2, 50-422 Wroc{\l}aw, Poland}

\date{\today}

\begin{abstract}
We report a comprehensive characterization of the magnetic, electrical transport, spectral, and thermal properties of single-crystalline Zintl-type material EuIn$_2$P$_2$. The compound crystallizes with a hexagonal unit cell (space group $P6_3/mmc$) and orders magnetically at $T_{\rm C}$~= 24~K with the Eu magnetic moments aligned ferromagnetically within the $ab$ plane but tilted alternately along the $c$–axis direction. The effective and saturation magnetic moments agree with the theoretical values expected for the Eu$^{2+}$ ion. For a range of several tens of kelvins above $\sim$40~K, the electrical transport of EuIn$_2$P$_2$ is dominated by short-range magnetic interactions. The temperature dependence of the electrical resistivity has been modelled in terms of variable-range hopping. Another indication of the latter scenario seems to be the observation for EuIn$_2$P$_2$ of a quadratic dependence of the negative magnetoresistance on the magnetic field strength and the scaled magnetization. The temperature dependence of Raman band position and FWHM, as well as phonon lifetime, confirms the presence of the distinctive regions observed in transport and magnetic studies.
\end{abstract}

\maketitle

\section{\label{sec:Intro}Introduction}

Eu-In-P Zintl phases have been the research subject for years because of the colossal magnetoresistance (CMR), widely known for manganites \cite{Jia2006}. The crux, however, lies in the mechanism responsible for the magnetic ordering and CMR in Eu-In-P systems. While in manganites, it is a double exchange or a superexchange, in the case of Eu-In-P, it is unclear \cite{Jia2006}. It has been suggested that with application of magnetic field the mobility edge is shifted away from the Fermi energy due to increasing magnetization, which provides additional charge carriers \cite{Pfu2008}.

EuIn$_2$P$_2$ crystallizes in a hexagonal unit cell (space group $P6_3/mmc$) and, as it is characteristic for Zintl compounds, the valence state is precisely defined, i.e. alternating layers of Eu$^{2+}$ and [In$_2$P$_2$]$^{2-}$ are present. Canted ferromagnetic ordering takes place at $T_{\rm C}$~= 24~K \cite{Pfu2008,Jia2006}. First-principles calculations indicate that the occupied 4$f$ states of Eu ions are located just below electron pockets at the M-point of the Brillouin zone \cite{Sin2012}.

The fascinating nature of EuIn$_2$P$_2$ stems also from its topological properties. \emph{Ab initio} simulation shows that it can be a ferromagnetic Weyl semimetal, in which Weyl point or Weyl nodal line occurs when the magnetic moments are oriented along $a$ or $c$ axis, respectively \cite{Sar2022}. Further electronic structure calculations have suggested that EuIn$_2$P$_2$ is a semiconductor with a band gap of 0.4~eV, and for comparison, a band gap in the isostructural EuIn$_2$As$_2$ compound results from the mixing of the valence band and the conduction band \cite{Shi2023}. On the other hand, other \emph{ab initio} calculations \cite{Sin2011} have suggested that the experimentally observed energy gap of 3.2~meV \cite{Jia2006} is too narrow and semiconducting behavior is an accidental coincidence, which results from contribution of different types of interband transitions. Holes are the dominant carriers in EuIn$_2$P$_2$ with the concentration of the order of 10$^{-19}$ cm$^{-3}$, as extracted from the Hall and Seebeck effect measurements \cite{Shi2023}.

In the current studies, we used complementary methods to verify the magnetic, transport, thermodynamic, and spectroscopic properties of the single crystalline EuIn$_2$P$_2$.

\section{\label{sec:Exp}Experimental methods}
\begin{figure*}[htbp]
\includegraphics[width=.7\textwidth]{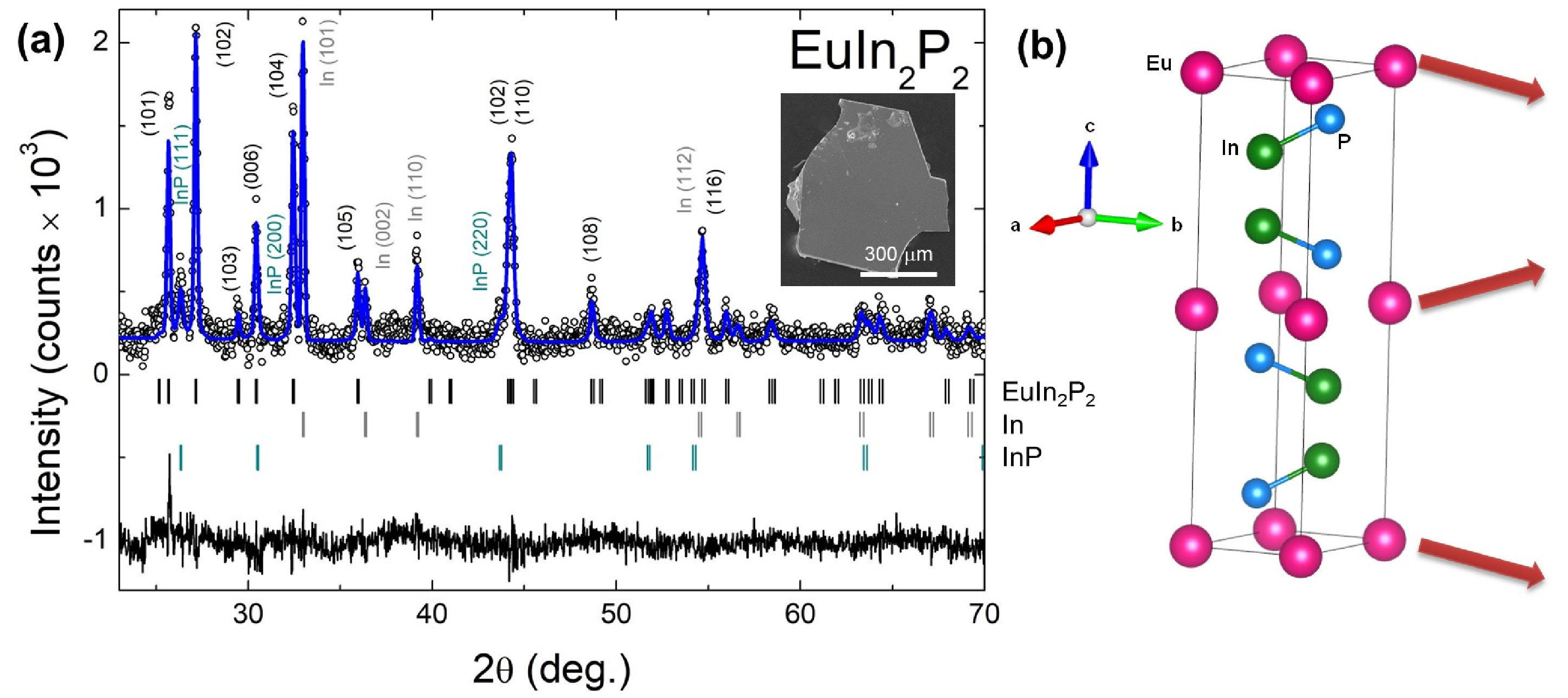}
\caption{\label{Fig1}
(a) Powder X-ray diffraction pattern with Rietveld refinement for EuIn$_2$P$_2$: experimental data (symbols), refined intensity (line), and their difference (lower curve). The top row of ticks represents Bragg markers for the hexagonal structure of the space group $P6_3/mmc$, the second row represents Bragg peak positions for pure In ($I$4/$mmm$), and third row represents Bragg peak positions for InP ($F$-43$m$). The most prominent peaks were described by Miller indices. The inset shows an electron microscope photograph of a selected impurity-free EuIn$_2$P$_2$ crystal. The visible flat surface presents the (001) surface.} (b) Crystal structure of EuIn$_2$P$_2$. The structure has a layered form composed of alternating anion layers of [In$_2$P$_2$]$^{2-}$ and cation layers of Eu$^{2+}$. Red arrows illustrate the orientation of the magnetic moments of the Eu layers as suggested in Refs. \cite{Pfu2008,Jia2006}.
\end{figure*}

EuIn$_2$P$_2$ crystals were synthesized by reaction of the pure elements with In playing the additional role of a flux. The elements in the ratio of 3:110:6 were placed in the ACP-CCS-5 Canfield Crucible Set (LCP Industrial Ceramics Inc.) sealed in an evacuated quartz tube. The reactants were kept at a temperature of 1100~$^{\circ}$C for 30 hours; then, the sample was cooled to 600$^{\circ}$C at a rate of 2~$^{\circ}$C/h. The residual In-flux was centrifuged at 600~$^{\circ}$C. The sample preparation process was adapted from the work \cite{Jia2006}.

The crystallographic structure was examined by X-ray diffraction (XRD) on powdered crystals using the Cu-K$_\alpha$ radiation. Within the space group $P6_3/mmc$ the Wyckoff positions of EuIn$_2$P$_2$ are: Eu 2$a$(0, 0, 0), In 4$f$(2/3, 1/3, u), and P 4$f$(1/3, 2/3, v) as illustrated in Fig.~\ref{Fig1}(b). The full-pattern Rietveld refinement of the XRD pattern measured at room temperature confirmed the hexagonal structure of the grown crystals and provided the lattice parameters values $a = b =$ 4.087(1)~{\AA} and $c$~= 17.622(4)~{\AA}. The values obtained agree with those published earlier \cite{Jia2006}.

XRD analysis was conducted on the ground EuIn$_2$P$_2$ crystals. These crystals had residual flux on their surface in the form of pure In and small amounts of InP compound, which was formed during the synthesis of EuIn$_2$P$_2$ single crystals. The largest single crystals of the EuIn$_2$P$_2$ phase, which were confirmed to be free of other phases through energy-dispersive X-ray spectroscopy (EDS), were selected for further investigation (see inset of Fig.~\ref{Fig1}(a)). The XRD detail are provided in the Supplemental Material \cite{suppl}.

The actual chemical composition of the crystals obtained was verified by energy-dispersive X-ray spectroscopy and the average value is Eu$_{1.01(3)}$In$_{2.09(6)}$P$_{1.90(6)}$. Measurements were collected from several locations on the surface of selected crystals.

The magnetic moment of the samples was measured in the temperature range of 1.9~– 400~K and magnetic fields up to 9~T using the Quantum Design Physical Property Measurement System (PPMS), which was equipped with a vibrating sample magnetometer (VSM).

The same PPMS platform was used for the heat capacity measurements using the relaxation technique and the two-$\tau$ method.

Electrical resistivity $\rho$ was also measured using PPMS. The data were collected in the temperature interval 2~– 300~K and in fields up to 9~T. Gold wires were fixed to the sample with silver-epoxy for the four-point transport measurements.

For analyzing spectral properties, a Jobin-Yvon HORIBA LabRAM HR 800 spectrometer with a 1024$\times$256 pixel LN-cooled CCD detector was utilized. Raman spectra from 30 to 3500~cm$^{-1}$ were recorded using He-Ne laser ($\lambda$~= 632.8~nm) as the excitation source, with a 50$\times$ microscope objective. The laser power was maintained below 1~mW to prevent sample degradation while the light was focused onto it. The achieved resolution exceeded 2~cm$^{-1}$. The calibration procedure was conducted utilizing the 520.7~cm$^{-1}$ silicon line. The samples were mounted in an Oxford Instruments Microstat cryostat cooled by liquid helium.

\section{\label{sec:Mag}Magnetic characterization}

\begin{figure*}
\includegraphics[width=0.7\textwidth]{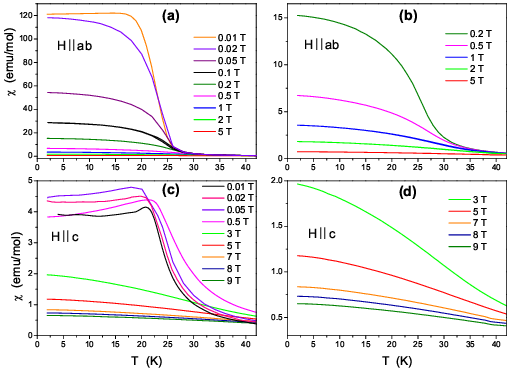}
\caption{\label{Fig2}
Magnetic susceptibility in low-temperature range for single crystalline EuIn$_2$P$_2$ measured for a set of values of the external magnetic field applied within the $ab$ plane (a,b), and along the $c$ axis (c,d). Panels (b) and (d) show magnification of the curves obtained at the highest magnetic fields.}
\end{figure*}

\begin{figure}
\includegraphics[width=1.0\linewidth]{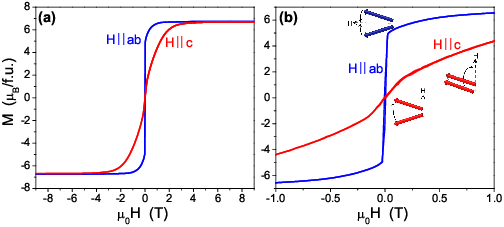}
\caption{\label{Fig3}
(a) Magnetization as a function of the magnetic field in the $ab$ plane and along the $c$ axis measured at 2~K. (b) Enlargement of the low-field part of the $M(H)$ curves. Thick arrows represent magnetizations of adjacent Eu layers.}
\end{figure}

Based on the temperature dependence of the magnetic susceptibility, it has been previously suggested that just above the ordering temperature, a short-range magnetic order may play a significant role in EuIn$_2$P$_2$ \cite{Jia2006,Hid2009}. We observed it also for single crystals of EuIn$_2$As$_2$ \cite{Tol2023}. Magnetic susceptibility ($\chi$) reveals also anisotropic behavior. For the magnetic field oriented within the $ab$ plane (Fig.~\ref{Fig2}(a,b)), $\chi$ vs. $T$ shows large values and typical ferromagnetic dependence below $T_{\rm C}$~= 24~K, while for the magnetic field aligned along the $c$-axis direction (Fig.~\ref{Fig2}(c,d)) the magnetic susceptibility is much smaller and small maxima are visible in low magnetic fields suggesting a contribution of antiferromagnetic (AFM) component. These maxima disappear in magnetic fields higher than the magnetic saturation field, $\mu_{\rm 0}H_{\rm sat}^{c} \approx$ 3~T, derived from the magnetization curve $M(H)$ measured in magnetic field applied along the $c$ axis (Fig.~\ref{Fig3}(a)).

\begin{figure}
\includegraphics[width=0.7\linewidth]{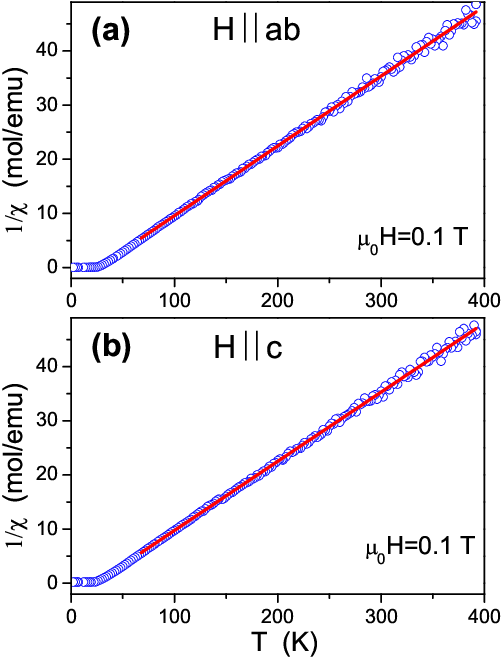}
\caption{\label{Fig4}
Inverse magnetic susceptibility of EuIn$_2$P$_2$ measured in the $ab$ plane (a) and along the $c$ axis (b) Curie-Weis fit is represented by solid red lines. See the main text for the fitting parameter values.}
\end{figure}

The slope change in $M(H)$ evident from Fig.~\ref{Fig3}(b) at $\mu_{\rm 0}H \sim$ 0.2~T for $H \parallel c$ originates from the spin-reorientation (SR) transition. In Fig.~\ref{Fig3}(b), thick arrows represent magnetizations of adjacent Eu layers (see also Fig.~\ref{Fig1}(b)), and thin black arrows indicate the direction of rotation of the magnetic moments in applied magnetic fields. In the initial state, the Eu magnetic moments are aligned ferromagnetically within the $ab$ plane, but tilted alternately towards the $c$ axis \cite{Jia2006}. The initial angle between the magnetizations of Eu layers is exaggerated in Fig.~\ref{Fig3} for better readability of the drawing. The tilt implies a small component of the magnetic moment $M_{\parallel}$ parallel to the $c$ axis but with alternating orientation, hence forming an AFM component with N{\'e}el vector along the $c$ axis. A small value of the initial $M_{\parallel}$ component explains a small magnetic field value (0.2~T) required to activate the SR process. For the magnetic field applied in the $ab$ plane, a coherent rotation of the magnetizations occurs, finally ending in saturation at
$\mu_{\rm 0}H_{\rm sat}^{ab} \approx$ 1~T. The rotation does not start precisely at zero field due to a tiny anisotropy within the plane, strain and/or domain structure.

Inverse magnetic susceptibility of EuIn$_2$P$_2$ obtained for magnetic field applied within the $ab$ plane and along the $c$ axis is plotted in Fig.~\ref{Fig4} after subtracting a small temperature-independent contribution, $\chi_0$, appearing due to the sample holder.

At temperatures larger than 80~K, which is well above the ordering temperature, $\chi^{-1}(T)$ was fitted with the Curie – Weiss (C-W) formula:

\begin{equation}
\chi^{-1} = \frac{3k(T-\theta_{\rm p})}{N_A \mu^2_{\rm eff}}.
\label{eq1}
\end{equation}
For both magnetic field orientations, consistent values of the fitting parameters were derived: the paramagnetic C-W temperature $\theta_{\rm p}$~= 26(1)~K and the effective magnetic moment $\mu_{\rm eff}$~= 7.87(2)~$\mu_{\rm B}$ for $H \parallel ab$, and $\theta_{\rm p}$~= 24(1)~K, $\mu_{\rm eff}$~= 7.91(2)~$\mu_{\rm B}$ for $H \parallel c$. For comparison, the theoretical value of $\mu_{\rm eff}$ for Eu$^{2+}$ ion is 7.94~$\mu_{\rm B}$. It also turns out that the obtained C-W temperature values are in good agreement with the magnetic ordering temperature.

\begin{figure}
\includegraphics{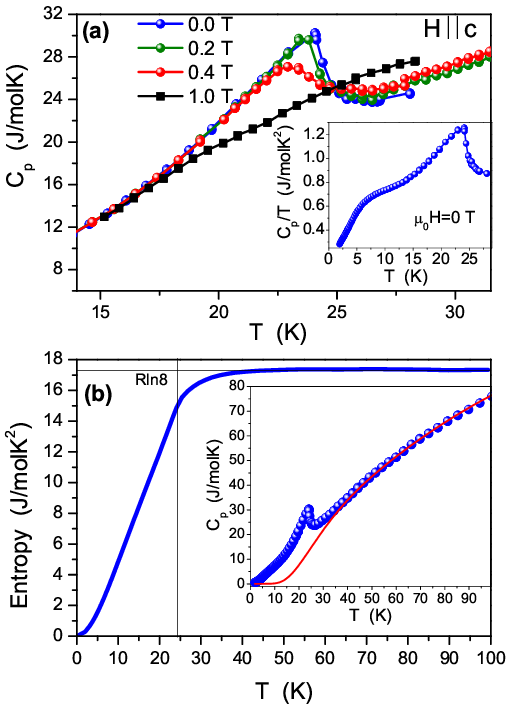}
\caption{\label{Fig5}
(a) Low-temperature specific heat measured for various magnetic fields applied along the $c$ direction. The peak shift towards lower temperatures with increasing magnetic field values corroborates the presence of the AFM component along the $c$ axis. Inset: Specific heat $C_{\rm p}(T)/T$ in the temperature range down to 2~K. (b) Temperature dependence of the magnetic entropy. Inset shows the fitting of the specific heat $C_{\rm p}(T)$ with the Debye-Einstein model (see Supplemental Material for details \cite{suppl}).}
\end{figure}

Further confirmation of the stacking of the ferromagnetic layers along the $c$ direction comes from measurements of the temperature-dependent specific heat $C_{\rm p}(T)$ at various magnetic fields applied along the $c$ axis (Fig.~\ref{Fig5}(a)). The shift of the $C_{\rm p}(T)$ peak towards lower temperatures with increasing magnetic field values corroborates the presence of the tiny AFM component along the $c$ axis. No other peak of $C_{\rm p}(T)$ was observed down to the temperature of 2~K. The inset of Fig.~\ref{Fig5}(a) presents $C_{\rm p}/T(T)$, which reveals a hump at about 6~K, which is typical of Eu$^{2+}$ systems. It stems from the Zeeman splitting of the $J$~= 7/2 multiplet. Fig.~\ref{Fig5}(b) shows the magnetic entropy as a function of temperature. The magnetic specific was obtained by subtracting the lattice contribution (inset) described by the Debye and Einstein models (see Supplemental Material for details \cite{suppl}). It seems from Fig.~\ref{Fig5}(b) that there is a small amount of entropy transferred above the ordering temperature.

\section{\label{sec:Resist}Electrical resistivity}

\begin{figure*}
\includegraphics{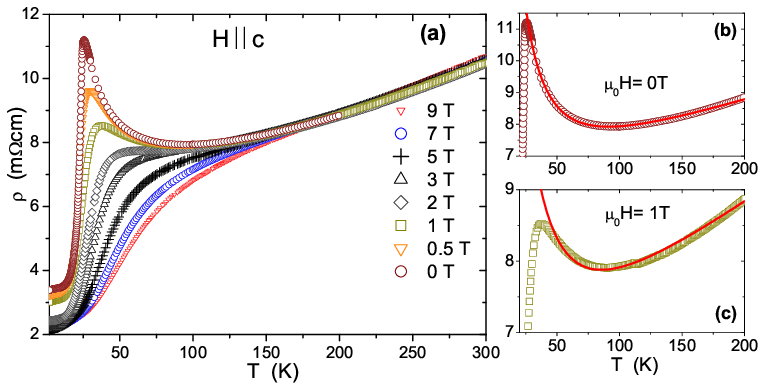}
\caption{\label{Fig6}
(a) Electrical resistivity of EuIn$_2$P$_2$ measured for electrical current within the $ab$ plane and increasing values of the magnetic field applied along the $c$ axis. Red curves in panels (b,c) represent the fitting with Eq.~\ref{eq2}.}
\end{figure*}

Temperature dependence of electrical resistivity of EuIn$_2$P$_2$ measured in zero magnetic field (Fig.~\ref{Fig6}) resembles results obtained previously for EuIn$_2$As$_2$ \cite{Tol2023}, i.e., above $T_{\rm C}$, it is dominated by short-range magnetic interactions \cite{Tol2023,Reg2020,Gon2022,Zha2020}. However, the absolute values of the resistivity are about twenty times higher. The alignment of the localized spins by the applied magnetic field leads to the visible resistance reduction, similarly to the results reported in Ref. \cite{Jia2006}.
\begin{figure}
\includegraphics[width=0.8\linewidth]{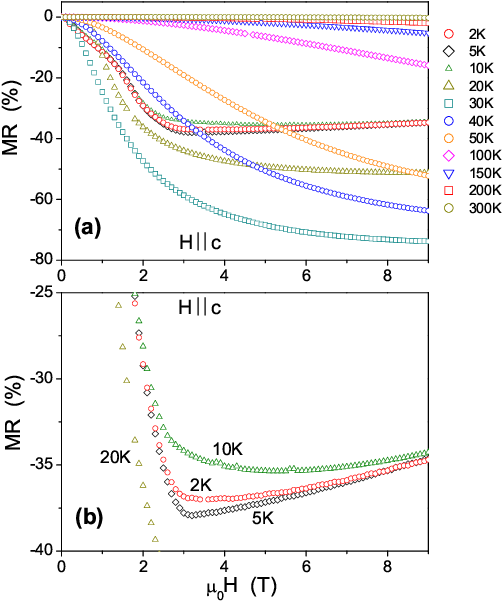}
\caption{\label{Fig7}
(a) Magnetoresistance isotherms for EuIn$_2$P$_2$ obtained with the magnetic field applied along the $c$ axis and electric current flowing in the $ab$ plane. (b) Enlargement of the low-temperature curves showing the change of the slope at the saturation magnetic field.}
\end{figure}
\begin{figure}
\includegraphics{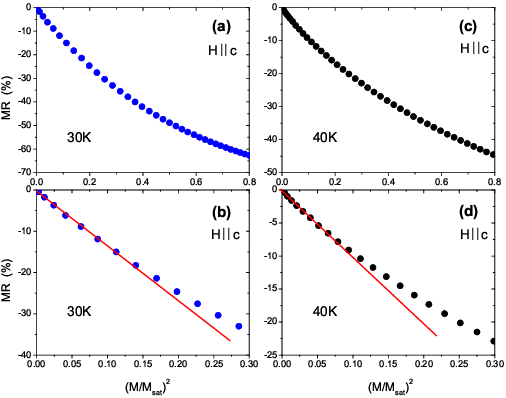}
\caption{\label{Fig8}
Magnetoresistance isotherms for EuIn$_2$P$_2$ at temperatures of 30~K and 40~K, i.e., above $T_{\rm C}$ (region of short-range order). Magnetic field was applied along the $c$ axis.}
\end{figure}

The $\rho(T)$ data for $T >$ 40~K were analyzed with the formula:

\begin{eqnarray}
\rho (T) = && a_1 + 4 a_2 \theta_{\rm D} \left( \frac{T}{\theta_{\rm R}} \right) ^5 \int_{0}^{\theta_{\rm R}/T} \frac{ x^5 dx }{ (e^x-1)(1-e^{-x})} \nonumber\\
&&+ a_3exp \left[ \left( \frac{T_{\rm M}}{T} \right) ^{1/4} \right],
\label{eq2}
\end{eqnarray}
where the first term accounts for temperature-independent scattering of electrons on structural defects and on disordered magnetic moments, the second term represents the Bloch-Gr\"uneissen function describing the electron-phonon scattering processes with $\theta_{\rm R}$ being a parameter related to the Debye temperature, and the last term is the Mott variable-range hopping (VRH) with $T_{\rm M}$ being a characteristic temperature related to the average energy level spacing \cite{Mot1968}. In Ref. \cite{Jia2006}, a simple semiconductor model has been used to account for the increase in resistivity with temperature decreasing from 60~K down to 29~K, providing a band gap of 3.2~meV, being two orders of magnitude smaller than expected for semiconductors. In our analyzes a good fit could be obtained only with the VRH. Figures~\ref{Fig6}(b,c) show a fit with Eq.~\ref{eq2} for $\mu_{\rm 0}H$ equal to 0~T and 1~T, and the obtained parameter values are listed in Table~\ref{tab1}. The parameters $\theta_{\rm R}$, $a_1$, and $a_2$, obtained for resistivity $\rho (T)$ measured in a zero magnetic field were fixed for $\rho (T)$ measured in non-zero magnetic fields because the first two terms of Eq.~\ref{eq2} should not be field-sensitive.

\begin{table}[b]
\caption{\label{tab1}%
Values of parameters obtained by fitting temperature dependence of the electrical resistivity of EuIn$_2$P$_2$ with Eq.~\ref{eq2} for magnetic field applied along the $c$-axis direction. The fixed parameters are: $\theta_{\rm R}$~= 289~K, $a_1$~= 5.75~m$\Omega$cm, and $a_2$~= 0.0127~m$\Omega$cm/K. The fitting accuracy of the numerical calculation was set to $\pm$1~K for temperatures and $\pm1\times10^{-5}$ for other parameters.}
\begin{ruledtabular}
\begin{tabular}{cccc}
\textrm{$\mu_{\rm 0}H$}&
\textrm{$T_{\rm M}$}&
\textrm{$T_{\rm M}$}&
\textrm{$a_3$}\\
\textrm{(T)}&
\textrm{(K)}&
\textrm{(eV)}&
\textrm{(m$\Omega$cm)}\\
\colrule
0 & 16078 & 1.385 & 0.0387\\
0.5 & 11099 & 0.956 & 0.0547\\
1.0 & 7917 & 0.682 & 0.0684\\
\end{tabular}
\end{ruledtabular}
\end{table}

As shown in Table~\ref{tab1}, the average energy gap, $T_{\rm M}$, is large for zero magnetic fields ($T_{\rm M}$~= 1.4~eV), but it decreases with the increasing value of the applied magnetic field. For comparison, the zero-field value of $T_{\rm M}$ for EuIn$_2$As$_2$ \cite{Tol2023} was equal to 0.25~eV. Magnetic field sensitivity of the activated resistivity above the ordering temperature may confirm the critical role of the short-range order. For the order introduced by the magnetic field, the probability of hoping increases, i.e., the gap decreases.

Qualitatively similar behavior of transport properties has also been observed in the EuSe system \cite{Sha1974,Hel1978,Kui1981,Cri2023}, where the activated behavior and a peak in the temperature dependence of resistivity have been found. The effect of the magnetic field also was the suppression of the resistivity peak, implying negative magnetoresistance. The properties of the EuSe semiconductor have been interpreted in frames of the bound magnetic polaron model \cite{Hel1978,Kui1981}. Similar scenario has been also claimed in Eu$_5$In$_2$Sb$_6$ \cite{Cri2023}.

The fundamental question concerns the type of interactions responsible for the magnetic order in EuIn$_2$P$_2$. As has been pointed out by Jiang et al. \cite{Jia2006}, the Eu-Eu distance (4.083~{\AA}) is too large to promote direct magnetic exchange interaction. Therefore, two other types of interaction to be considered are the double-exchange (DE) and the Bloembergen-Rowland interaction \cite{Pfu2008,Jia2006}. DE has been suggested in some Eu-based compounds, e.g., hexaborides Eu$_{1-x}$Ca$_x$B$_6$ \cite{Wig2002,Per2004}. For EuIn$_2$P$_2$ below the ordering temperature, the long-range magnetic order is probably governed by the Bloembergen-Rowland interaction.

\begin{figure}
\includegraphics[width=0.8\linewidth]{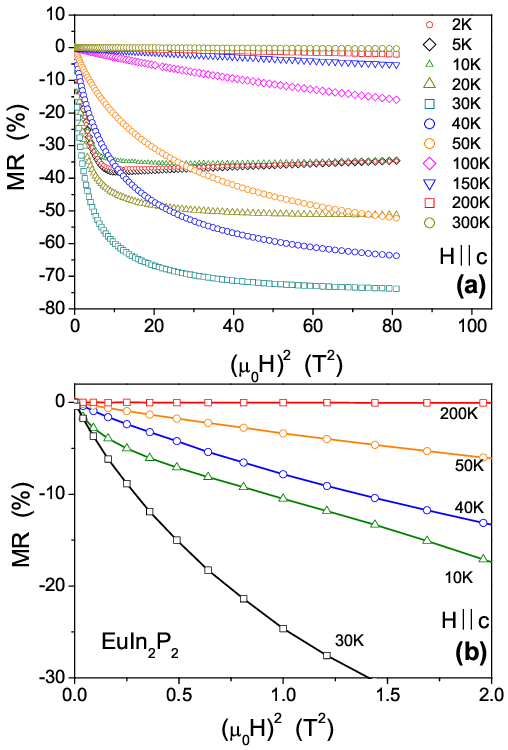}
\caption{\label{Fig9}
(a) Magnetoresistance isotherms of EuIn$_2$P$_2$ as a function of $(\mu_0 H)^2$. (b) Enlargement of the low magnetic field range. Magnetic field was applied in the $c$-axis direction.}
\end{figure}
\begin{figure}
\includegraphics{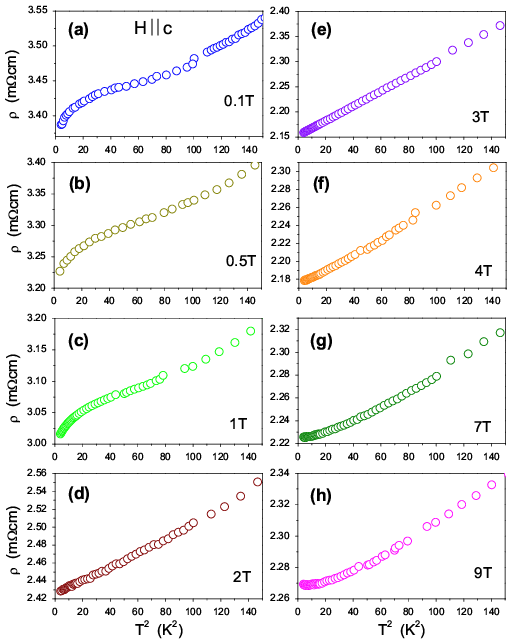}
\caption{\label{Fig10}
Low-temperature resistivity of EuIn$_2$P$_2$ as a function of $T^2$ at various values of the magnetic field applied in the $c$-axis direction. Linear behavior is revealed for curves measured at around 3~T.}
\end{figure}

Furthermore, if magnetic polarons are relevant, magnetoresistance is expected to follow the Majumdar-Littlewood (ML) model \cite{Zha2020,Maj1998}, which implies that at low magnetic fields, magnetoresistance fulfills the relation:

\begin{equation}
{\rm MR} = \frac{\left( \rho (H)-\rho (0)\right)}{\rho(0)} \propto  \left( \frac{M}{M_{\rm sat}} \right)^2.
\label{eq3}
\end{equation}

Figure~\ref{Fig7} shows magnetic-field dependence of MR. Both the MR behavior with increasing magnetic field and with increasing temperature reflect the ferromagnetic nature of the crystal studied. The ML model states that the relationship of Eq.~\ref{eq3} should hold for small magnetic field values. This dependence is plotted in Fig.~\ref{Fig8} for the temperatures above $T_{\rm C}$, i.e., in the region where the short-range order is postulated for EuIn$_2$P$_2$. The linear behavior can be seen for small values of $(M/M_{\rm sat})^2$ corresponding to small magnetic field values, as expected from the ML model.

Next, negative MR (NMR) and its quadratic dependence on the magnetic field may be another indication of the VRH mechanism in the region of short-range interactions above $T_{\rm C}$ \cite{Syb2012}. To highlight the quadratic behavior, Fig.~\ref{Fig9} shows MR as a function of $(\mu_{\rm 0}H)^2$. The quadratic dependence is observed already in the region of short-range magnetic order, before the clear metallic behavior starting at about 150~K, where the linear dependence of MR on $(\mu_{\rm 0}H)^2$ is typically expected.

Interestingly, in Fig.~\ref{Fig7}(b), below the ordering temperature and for magnetic field larger than the saturation field, a change in slope of MR($(\mu_{\rm 0}H)$) dependence is visible, which may be a signature of a topological phase transition. Such a possibility has been suggested by Sarkar et al. \cite{Sar2022} based on first-principles calculations and symmetry analysis. Those computations indicated that EuIn$_2$P$_2$ can be a Weyl semimetal when saturating magnetic field is applied in the $ab$ plane and a nodal-line semimetal for magnetic field saturating magnetization in the $c$-axis direction. The nodal-line semimetal should show a quadratic dependence of MR on the magnetic field for low magnetic field values, followed by linear non-saturating magnetoresistance at higher magnetic fields \cite{Gao2023,Wan2012a,Wan2012b,Ren2018}.
In the current case, studies in much higher magnetic fields are necessary to confirm such a scenario. However, it is puzzling that at low temperatures resistivity exhibits perfectly linear dependence on $T^2$ for magnetic field values close to the magnetic saturation $\mu_{\rm 0}H_{\rm sat}^{c} \approx$3~T (Fig.~\ref{Fig10}). It may suggest an increased contribution of the electron-electron scattering and may be related to the activation of the nodal state with $\mathbb{Z}_4$~= 1 (see Fig.~\ref{Fig4}(g) in Ref.~\cite{Sar2022}). Moreover, in the VRH regime, the density of states is assumed to be constant because electron-electron interactions are neglected \cite{Lim2021} and this explains why for magnetic field values higher than about 2~T, the maximum in the $\rho(T)$ dependence (Fig.~\ref{Fig6}) is damped - VRH is suppressed and the increased role of electron-electron interactions is revealed (Fig.~\ref{Fig10}). The damping of the resistivity peak occurs because magnetic field reduces the magnetic disorder, which implies decreasing scattering probability within the VRH mechanism.

\section{\label{sec:Ram}Raman spectroscopy}
\begin{figure*}
\includegraphics{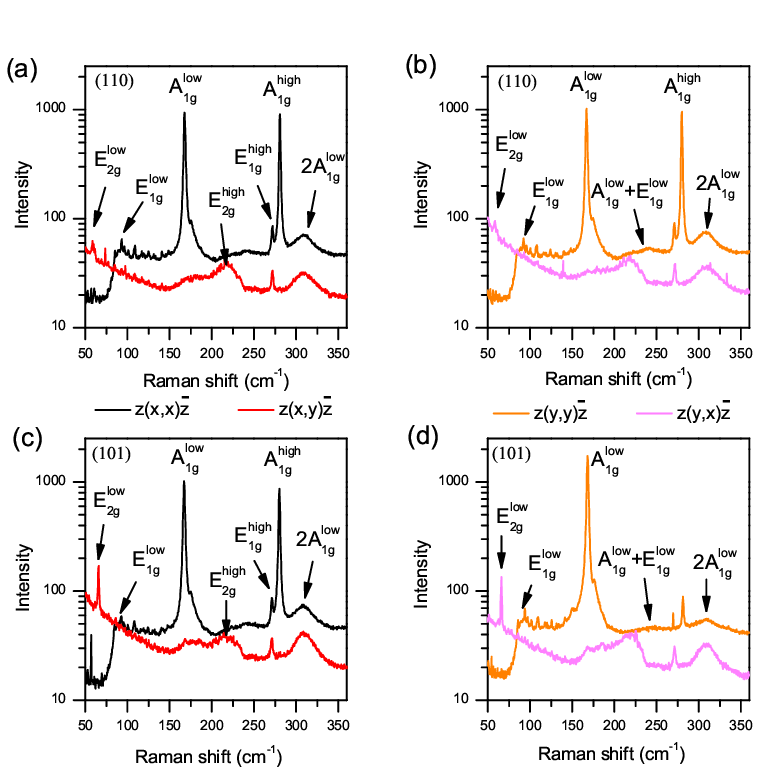}
\caption{\label{Fig11}
Room temperature Raman spectra of EuIn$_2$P$_2$ in $z(xx)\bar{z}$, $z(xy)\bar{z}$, $z(yy)\bar{z}$, $z(yx)\bar{z}$ geometry performed for two natural faces developed on the single crystal: (110) (a,b) and (101) (c,d). Note: the ordinate axis is on a logarithmic scale.}
\end{figure*}
\begin{figure}
\includegraphics[width=0.7\linewidth]{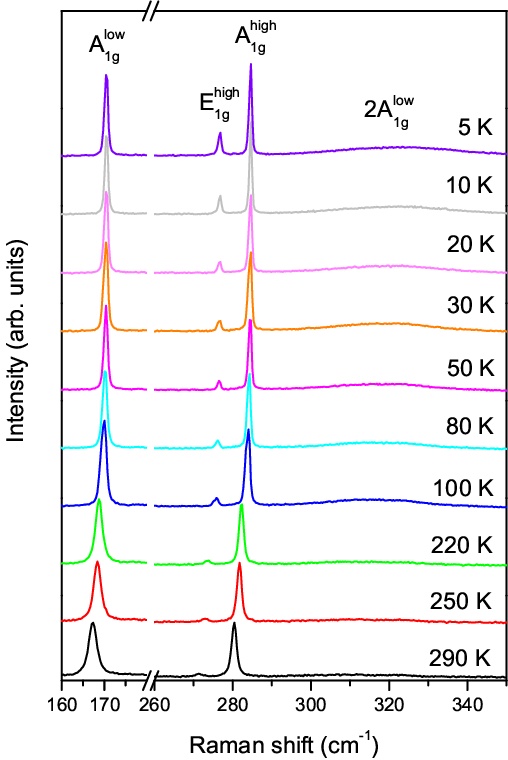}
\caption{\label{Fig12}
 Raman spectra of EuIn$_2$P$_2$ performed for the growth face developed of the single crystal (110) from 5 to 290~K. Note: the electric vector in the excitation beam was polarized parallel to the [100] direction.}
\end{figure}

EuIn$_2$P$_2$ contains two formula units per hexagonal unit cell with the space group $D_{6h}^4$ ($P6_3/mmc$). Since there are four In atoms, four P atoms, and two Eu atoms, it is immediately seen by inspection of the Space-group Table No. 194 \cite{InterTab1952} that the indium and phosphorus atoms each must occupy $fC_{3v}^d$(4) sites and the europium atoms must be on $aD_{3d}$(2) sites. To analyze the Raman spectra of EuIn$_2$P$_2$ single crystals, we have utilized the nuclear site group analysis method. This method involves referring to the tables provided in Reference \cite{Rou1981}, which contain information about the irreducible representations that arise from occupying different sites within the $D_{\rm 6h}$ space group. Specifically, Table 27B in this reference is used for this purpose. Each nucleus on $C_{3v}^d$ site will contribute $A_{\rm 1g}+A_{\rm 2u}+B_{\rm 1g}+B_{\rm 2u}+E_{\rm 1g}+E_{\rm 1u}+E_{\rm 2g}+E_{\rm 2u}$ modes and nucleus on $D_{3d}$ site will contribute $A_{\rm 2u}+B_{\rm 2u}+E_{\rm 1u}+E_{\rm 2u}$, giving a total of $2A_{\rm 1g}+3A_{\rm 2u}+2B_{\rm 1g}+3B_{\rm 2u}+2E_{\rm 1g}+3E_{\rm 1u}+2E_{\rm 2g}+3E_{\rm 2u}$. The acoustic modes are $A_{\rm 2u}+E_{\rm 1u}$, as indicated in the Character table of $D_{\rm 6h}$. The optical-phonons belong to the following irreducible representations: $\Gamma_{\rm opt} = 2A_{\rm 1g}+2A_{\rm 2u}+2B_{\rm 1g}+3B_{\rm 2u}+2E_{\rm 1g}+2E_{\rm 1u}+2E_{\rm 2g}+3E_{\rm 2u}$. The $2B_{\rm 1g}+3B_{\rm 2u}+3E_{\rm 2u}$ are silent modes, $2A_{\rm 2u}+2E_{\rm 1u}$ are IR active, and $2A_{\rm 1g}+2E_{\rm 1g}+2E_{\rm 2g}$ are Raman active modes.

The IR active modes split into transverse and longitudinal waves. Only transverse phonons can interact with infrared radiation; thus, this will not impact the IR absorption spectrum. Raman active phonons are not simultaneously IR active, so no frequency splitting between the transverse and longitudinal waves is present in Raman spectra. Using these theoretical predictions, one can anticipate finding two intense lines with $A_{\rm 1g}$ symmetry in the Raman spectra and much weaker lines corresponding to doubly degenerate modes. These Raman active modes are related to the dynamics of anion layers of [In$_2$P$_2$]$^{2-}$ because the indium and phosphorus atoms occupy $fC_{3v}^d$(4) sites. Europium atoms occupying $aD_{3d}$(2) sites yield active modes only in the infrared spectrum.
\begin{figure*}
\includegraphics{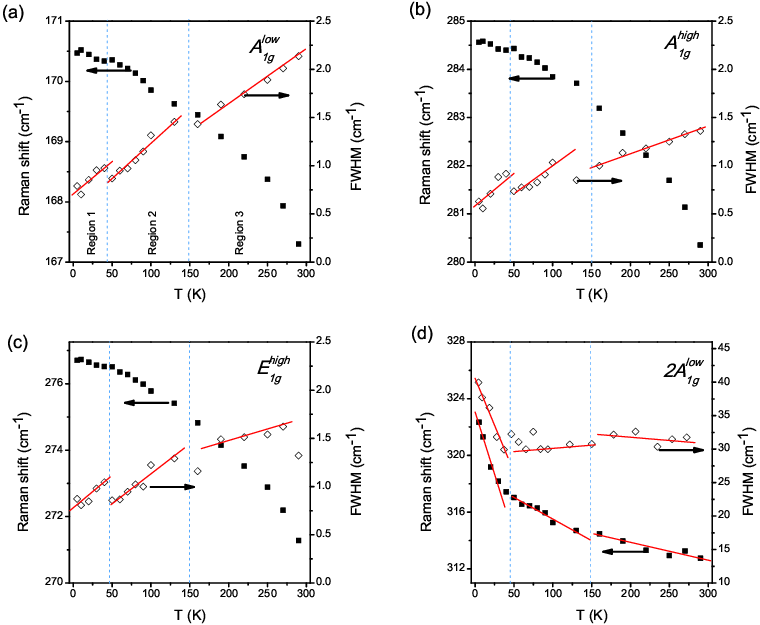}
\caption{\label{Fig13}
Temperature-dependent Raman shifts and FWHM of mode $A_{1g}^{low}$, $E_{1g}^{high}$, $A_{1g}^{high}$, 2$A_{1g}^{low}$. Red solid lines are guides for the eyes.}
\end{figure*}
\begin{figure}
\includegraphics{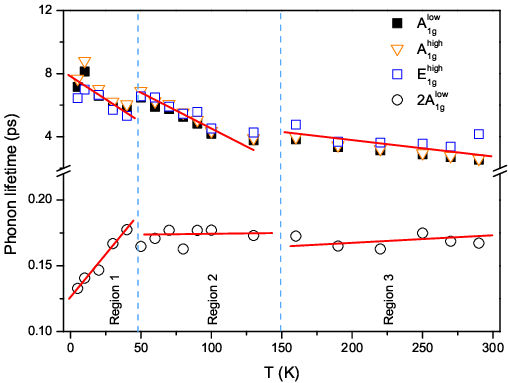}
\caption{\label{Fig14}
Comparative plots of temperature dependence of phonon lifetime of $A_{1g}^{low}$, $E_{1g}^{high}$, $A_{1g}^{high}$, and 2$A_{1g}^{low}$ modes. Red solid lines are guides for the eyes.}
\end{figure}

The detailed spectroscopic investigations we have performed for an oriented crystal of EuIn$_2$P$_2$. Depolarization of Raman spectra analysis was performed for two growth faces developed of the single crystal: the biggest crystallographic natural plane (110) and (101) (see Fig.~\ref{Fig11}). To simplify the description of the direction of the measurement, we introduce a Cartesian system with X, Y, and Z axes. The X-axis in the first case was parallel to the [100] direction and in the second to [001]. There are four different excitation and scattering polarization combinations possible: $z(xx)\bar{z}$, $z(xy)\bar{z}$, $z(yy)\bar{z}$, $z(yx)\bar{z}$. Parallel and cross polarizations, i.e., $z(xx)\bar{z}$, $z(xy)\bar{z}$ in Porto notation, probe the diagonal and off-diagonal elements of the Raman tensor, respectively. $E$ modes can be observable in both $z(xx)\bar{z}$, $z(xy)\bar{z}$ geometry, while $A$ is only visible in $z(xx)\bar{z}$ geometry. Based on the depolarization coefficients of the Raman bands displayed in Fig.~\ref{Fig11}, the assignment of the Raman active bands was proposed. Polarized peaks at 167 and 280~cm$^{-1}$ are related to $A_{1g}^{low}$ and $A_{1g}^{high}$ modes, respectively; low and high refers to the frequency. Figure 11 demonstrates that the mod $A_{1g}^{high}$ represents the stretching of two In atoms moving relative to one another along the $c$-axis. In Fig.~\ref{Fig11}(d), the electric vector in the excitation beam is polarized perpendicular to the [001] direction, which is also perpendicular to the In–In bonds. In this case, the band $A_{1g}^{high}$ cannot be seen.

The 93, 271~cm$^{-1}$ and 59, 217~cm$^{-1}$ bands correspond to $E_{\rm 1g}$ and $E_{\rm 2g}$ modes, respectively. The $E_{1g}^{high}$ mode (271~cm$^{-1}$) represents the symmetric bending of P atoms. The broad feature at around 312 cm$^{-1}$ (0.04~eV) may be attributable to the overtone scattering of the  $A_{1g}^{low}$ phonon, which is broadened due to the significant downward dispersion of the phonon branch \cite{Sin2012}. The broad, weak feature at about 240~cm$^{-1}$ may be assigned to two-phonon excitation of $A_{1g}^{low}$+$E_{1g}^{low}$.

Raman linewidth analysis vs temperature or pressure can reveal the interaction of phonons with the magnetic system (spin-phonon coupling), electronic system (electron-phonon coupling), and anharmonic interaction (phonon-phonon coupling) \cite{Du2019,Akr2012,Raj2019}. We performed temperature-dependent Raman measurements for the biggest crystallographic natural plane (110); the electric vector in the excitation beam was polarized parallel to the [100] direction (see Fig.~\ref{Fig12}).

The temperature dependencies of the Raman shift and full width at half maximum (FWHM) for the strongest modes: $A_{1g}^{low}$, $E_{1g}^{high}$, $A_{1g}^{high}$, 2$A_{1g}^{low}$ are shown in Fig.~\ref{Fig13}. The temperature dependence of the first-order Raman modes' wavenumber and FWHM are comparable, as shown in Fig.~\ref{Fig13}(a-c). As temperature was decreased from 290 to 5~K, the peak position of the $A_{1g}^{low}$, $A_{1g}^{high}$, and $E_{1g}^{high}$ was shifted from 167.3 to 170.5~cm$^{-1}$ ($\Delta\nu$ = 3.2~cm$^{-1}$), from 280.4 to 284.6~cm$^{-1}$ ($\Delta\nu$~= 4.2~cm$^{-1}$), and from 271.3 to 276.7~cm$^{-1}$ ($\Delta\nu$~= 5.4~cm$^{-1}$), respectively. The hardening and narrowing of the first-order Raman modes are due to the anharmonic phonon coupling. The variations in the 2$A_{1g}^{low}$ band position ($\Delta\nu$~= 10~cm$^{-1}$) and half-width ($\Delta$FWHM~= 8~cm$^{-1}$) as temperature increases (Fig.~\ref{Fig13}(d)) are significantly greater than those observed for the modes mentioned above. In this case, anharmonic interaction is more pronounced, which implies stronger phonon-phonon coupling. In the temperature dependence of these parameters, one can see an anomaly in the region of 24~K, which agrees with the observed by us temperature dependence of electrical resistivity. Red solid lines in Fig.~\ref{Fig13} are guides for the eyes to discriminate roughly the characteristic regions resulting from the magnetic and transport results: 1 – region below and around the ordering temperature, 2 – short-order correlations, 3 – pure paramagnetic order. It should be noted that the slight change in modes behavior at around 150~K coincides with the temperature at which the effect of magnetic field on the resistivity ceases (see Fig.~\ref{Fig6}). It is evident that the Raman spectroscopy data corroborate observations derived from the magnetic and transport measurements.

The main phonon broadening mechanism is the anharmonic decay into two, three, or more Brillouin zone phonons while conserving wavevector and energy. The identification of the most probable phonon decay channel can be accomplished through an examination of the temperature dependence of the phonon FWHM \cite{Pom2005,Iqb2019}. For EuIn$_2$P$_2$, in-depth comprehension of the decay mechanism necessitates an examination of the phonon dispersion curves \cite{Sin2012}. From the Raman peaks, the phonon lifetime $\tau$ can be estimated using the relationship, 1/$\tau$~= $\Gamma /\hbar$, where $\Gamma$ represents the FWHM of a peak in the unit of cm$^{-1}$, $\hbar$ is the Plank constant, and its value is 5.3 $\times$
 10$^{-12}$ cm$^{-1}$s \cite{Pom2005}. When analyzing the bandwidth, we considered the spectrometer broadening width, which was determined by accurately measuring the plasma line at the same spectrometer slit width as in the Raman experiments.

The phonon lifetimes estimated for $A_{1g}^{low}$, $E_{1g}^{high}$, $A_{1g}^{high}$, 2$A_{1g}^{low}$ are shown in Fig.~\ref{Fig14}. It can be seen that with increased temperature, the phonon lifetime of first-order Raman bands decreases. At a temperature of 50~K, the phonon lifetimes for the $A_{1g}^{low}$, $E_{1g}^{high}$, and $A_{1g}^{high}$ were determined to be 6.5, 6.9, and 6.6~ps, respectively. This long phonon lifetime suggests that 'hot phonon' effects will play a role in EuIn$_2$P$_2$ for carrier relaxation. For materials with a sufficiently long optical phonon lifetime, 'hot phonon' effects slow down carrier relaxation and reduce the drift velocity \cite{Pom2005, Ye2001}.
The hot phonon effects can be especially effective at moderate carrier densities (around 10$^{19}$ cm$^{-3}$), as was shown, e.g., for halide perovskites \cite{Fu2017}. For EuIn$_2$P$_2$, based on Hall effect measurements, we observed carrier densities in a similar range of values ($\thicksim 10^{19}$ cm$^{-3}$).
Therefore, the reduction of charge mobility is due to enhanced electron–phonon scattering \cite{Zan2004}. This effect influences the electrical properties of the material being studied. As the temperature decreases from about 80~K to approximately 24~K, there is an increase in electrical resistance. At low temperatures, below 24~K, the effect of the overtone vibration $A_{1g}^{low}$, causes the electrical resistivity to begin to decrease. As the temperature drops from 24~K to 5~K, the phonon lifetime of 2$A_{1g}^{low}$ decreases from 0.18~ps to 0.13~ps. The observed temperature dependency of phonon lifetime and its relatively low value suggest that this contribution can enhance charge mobility in low-temperature regions.

\section{\label{sec:Con}Conclusions}

The main observations of the current studies on EuIn$_2$P$_2$ are as follows:

\begin{enumerate}[noitemsep,topsep=0pt]
    \item Due to the large Eu-Eu distance the long-range magnetic ordering below ($T < T_{\rm C}$~= 24~K) is enabled most likely by the Bloembergen-Rowland interaction, whereas the short-range order is relevant for the range of several tens of kelvins above $\sim$40~K and is confirmed by:
       \begin{itemize}[noitemsep,topsep=0pt]
       \item Variable range hopping contribution to the temperature dependence of resistivity. The gap value (0.5~eV for $H~> 0$ and 1.3~eV for $H$~= 0) agrees with previous \emph{ab~initio} calculations.
       \item The dependence MR $\sim H^2$ is observed already in the region of short-range magnetic order, before the region with clear metallic behavior (above about 150~K), where the quadratic relationship is typically expected.
       \item MR $\sim$ $(M/M_{\rm sat})^2$ for small magnetic field values.
       \item Temperature dependence of the Raman shifts, FWHM, and phonon lifetime of the identified Raman modes correspond well to the characteristic regions observed in the transport and magnetic measurements.
       \end{itemize}
   \item Change of of the MR($H$) dependence slop from negative to positive after crossing the saturation field $\mu_{\rm 0}H_{\rm sat}^{c}$ (below $T_{\rm C}$) may be assigned to transformation to a nodal-line semimetal state. An additional argument confirming such a scenario may be the linear dependence of the resistance on $T^2$ at the magnetic saturation field because it implies enhanced electron-electron scattering contribution. The appropriate magnetic field probably shifts the occupied 4$f$ states of Eu ions to fit them with the electron pockets at the M-point of the Brillouin zone. It also coincides with closing the gap above $T_{\rm C}$, where short-range order exists (for 0 or small magnetic fields) and $\rho(T)$ is well described by the VRH mechanism. VRH stops to be relevant when e-e interactions become essential (when magnetic field - induced order closes the VRH gap).
    \item The Raman spectroscopy revealed that the transport properties of EuIn$_2$P$_2$ above $T_{\rm C}$ are governed by 'hot phonon' effects reducing the charge mobility due to enhanced electron–phonon scattering.
\end{enumerate}

\begin{acknowledgments}
This study was supported by the National Science Centre (Poland) under grant 2021/41/B/ST3/01141.
\end{acknowledgments}

\nocite{*}

\bibliography{EuIn2P2-paper}

\end{document}